\documentclass[preprint]{jpsj2}

\def\runtitle{A Toy Model of  Flying Snake's Glide}
\def\runauthor{Matsumura and Taguchi}

\title{%
A Toy Model of  Flying Snake's Glide}

\author{%
$^1$ Koji  Matsumura\thanks{E-mail address: matsu@phys.chuo-u.ac.jp} and
$^{1,2}$Y-h. Taguchi\thanks{E-mail address: tag@granular.com}
}

\inst{%
$^1$ Department of Physics, Faculty of Science and Technology,
Chuo University, 1-13-27 Kasaga, Bunkyo-ku, Tokyo 112-8551.\\
$^2$ Institute for Science and Technology,
Chuo University, 1-13-27 Kasaga, Bunkyo-ku, Tokyo 112-8551.
}

\recdate{\today}

\abst{We have developed a toy model of flying snake's glide
[J.J. Socha,  {\it Nature} {\bf 418}  (2002) 603.] by
modifying a model for a falling paper.
We have found that asymmetric oscillation is a key about
why snake can glide.
Further investigation for snake's glide
will provide us details about how it can glide without a wing.
}

\kword{%
flying snake, glide, toy model, a falling paper
}

\begin{document}
\sloppy
\maketitle

\section{Introduction}
Biological fluid mechanics\cite{bio_fluid,bio_fluid2} is recently an interesting field.
It is unclear why dolphins or tuna fishes can swim so rapidly
and why small insects like flies or bees can get large enough lift force.
However a flyingsnake\cite{snake} is stranger than them 
in the light of fluid mechanics.
Although it does not have anything like wings, 
it can glide with constant falling speed.
This means it can produce at least some lift force which cancels
downward gravity.
Here we have shown how it can glide without wings using the toy model
\cite{Tanabe-Kaneko} which is originally proposed for 
explaining the
behaviour of a falling paper.
A falling paper exhibits several behaviour ranging from
a simple falling to a periodic rotation\cite{paper}.
However, there are no directional gliding motions 
which the flyingsnake can have.
We have found asymmetric oscillating motion can
induce a gliding motion. Probably this is the reason
why the flyingsnake must oscillate when it glides.

When the flying snake would like to glide\cite{snake}, it jumps up from some
higher location. After some short transient time period,
it starts to glide. 
While gliding, it shapes on horizontal plane a S letter. 
Furthermore head-to-tail distance oscillates periodically.
It seems to get lift force without doing anything special
other than that.
It is the rather difficult task to understand by fluid mechanics
this mechanism.
Fluid around a falling snake is
disturbed violently and its spatio-temporal patterns are very complicated,
probably something like that called turbulence appears.
It is difficult not only to understand such complicated spatio-temporal
patterns but also to simulate them numerically.

\section{A Falling Paper Model and its Modification for
Flying Snake} 
Tanabe-Kaneko model\cite{Tanabe-Kaneko} can
reproduce qualitatively the behaviour of a falling paper;
chaotic or periodic rotations,
chaotic or periodic flutterings and a simple falling,
although there are some criticisms to this model\cite{comments}.
In Tanebe-Kaneko model, a falling paper has only three state variables;
a horizontal velocity, a vertical velocity, 
and an angle of inclination of paper.
This means, a falling paper is modeled as a line segment which
goes down in vertical plane. 
Thus its motion is restricted to be
within two dimensions.
External forces are a gravity, a viscosity force and a lift.
Although the later two forces are complicated functions of the velocity field
of the surrounding fluid, Tanabe and Kaneko
assume they depend upon only a horizontal 
velocity, a vertical velocity and
an angle of inclination of paper.
In this study we employ this model in order to
describe a flying snake's gliding motion.

In order to model snake's gliding with this model,
we have assumed that
\begin{enumerate}
\item 
Neither viscosity force nor lift  changes
due to oscillations. Viscosity force is mainly dependent upon
level cross-section area on which snake shapes S letter.
This area does not change due to oscillation.
Lift  mainly depends upon snake's volume, which
is conserved during oscillation.
\item 
Inertia moment is strongly dependent upon oscillation because
head-to-tail distance is proportional to its gyration radius whose
squared value is proportional to inertia moment.
\end{enumerate}
Except for those mentioned above, snake is regarded as a falling paper.

When snake oscillates, the distance $d$ between head and tail has
 time dependence as $d(t)$.
Other than making inertia moment dependent
upon time,
we do not modify Tanabe-Kaneko model at all. 
As a result, our model for flying snake is\cite{num}
\begin{equation}
\begin{split}
\dot{u}&=-(k_\perp \sin^2\theta+k_\parallel \cos^2\theta)u
 +(k_\perp-k_\parallel) \sin \theta \cos \theta v 
  \mp \pi \rho V^2 \cos(\alpha +\theta)\cos \alpha, \\
\dot{v}&=(k_\perp - k_\parallel)\sin \theta \cos \theta u
-(k_\perp \cos^2 \theta + k_\parallel \sin^2 \theta)v 
\pm \pi \rho V^2\cos(\alpha +\theta)\sin \alpha -g, \\
\dot{\omega}&=-k_\perp \omega -(3\pi \rho V^2/d(t))
\cos(\alpha+\theta)\sin(\alpha + \theta), \label{eq1}
\end{split}
\end{equation}
where fluid density ratio $\rho$ is $\rho_f \ell_c/m_p$,
$\rho_f$ is fluid density, $m_p$ and $\ell_c$ are mass and
horizontal size of snake respectively.
As mentioned above, $\ell_c$ does not change with oscillation but
takes constant value.
$u$ and $v$ are horizontal velocity and vertical one, $\omega$ is 
the angular velocity on the vertical plane. $k_\perp$ ($k_\parallel$) is the friction coefficients along the perpendicular
(parallel) direction to 
a plane on which a snake shapes a S letter.
$\theta$ is incline angle
of this plane. $\alpha \equiv \arctan(u/v)$ and $V= \sqrt{u^2+v^2}$.

\section{Results}
We have tried to find $d(t)$ which 
induces the
gliding motion which
Tanabe-Kaneko original model lacks. 
First of all, we employ simple harmonic motion as $d(t)$.
It cannot induce gliding motions at all,
although we check any possibilities 
that original model exhibits
any motions.
Thus we tried other oscillating motions. After several trials and errors,
we found three cases which can produce gliding motion.
These three  $d(t)$s are
$$
d(t) = \left\{
\begin{array}{l}
0.3(1+\sin \omega_0 t)^{2}+0.7, \label{l(t)1} \\
\exp(-1-1.3 \sin\omega_0 t), \label{l(t)1}\\
1/[1+0.5\sin\omega_0 t]. \label{l(t)2}
\end{array}
\right .
$$
In the case of all of these three,
$d(t)$ oscillates around 1.0.
When $d(t)$ takes the constant values, 
this corresponds to original model of a falling paper.
Also, as long as we tried, gliding motion can take place
only when 
$d(t)$ oscillates around the value 
with which a periodic fluttering occurs in the original model.
\begin{figure}
\includegraphics[width=80mm]{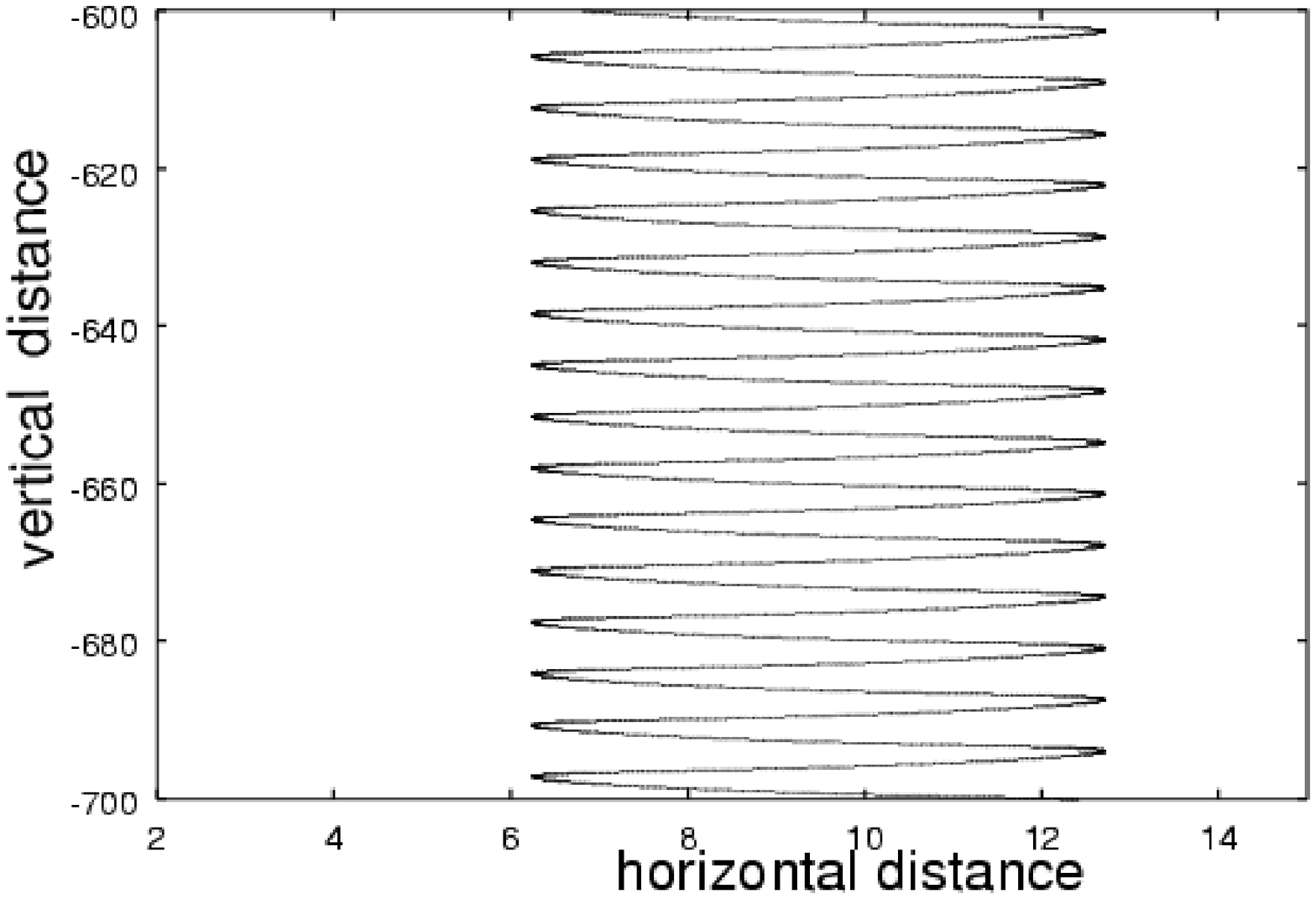}
\includegraphics[width=80mm]{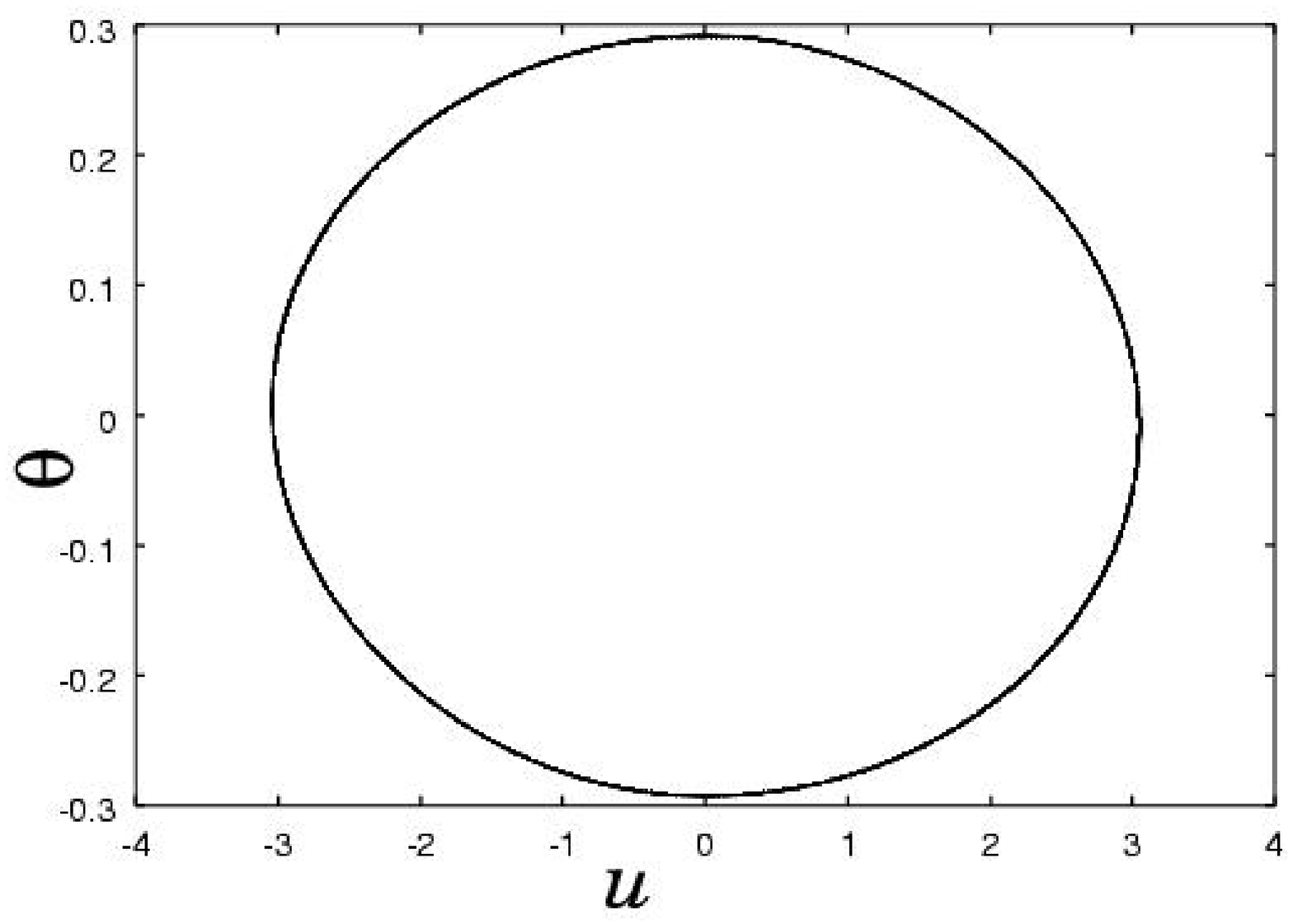}
\caption{Left: trajectory of periodic fluttering of  a falling paper
reproduced by Tanabe-Kaneko model ($d(t)=1$).  
Right: $\theta(t)$ as a function of $u(t)$.
Their relation is the same as that
between the velocity and the coordinates 
observed when harmonic motion takes place.}
\label{fig1}
\end{figure}
In Figure \ref{fig1}, we show periodic fluttering of a falling paper.
We also show $\theta(t)$ as a function of $u(t)$.
Due to the symmetric motion, their relation is also
symmetric and is nearly equal to that
between the velocity and the coordinates
observed when a harmonic motion takes place.

Now in order to simulate snake case, 
we have to consider the time dependence of $d(t)$.
\begin{figure}
\includegraphics[width=86mm]{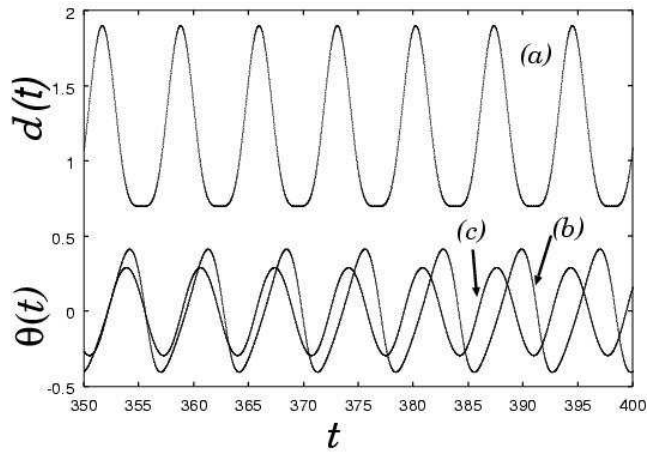}
\caption{(a) $d(t)=0.3(1+\sin \omega_0 t)^{2}+0.7$, $\omega_0=0.9$ 
(b) $\theta(t)$ when $d(t)=1$. 
Its time dependence is almost
sinusoidal because paper flutters 
periodically.
(c) $\theta(t)$ when $d(t)$ varies as shown in (a).
Due to the oscillation, period of fluttering becomes close to
that of oscillation.}
\label{fig2}
\end{figure}
In Figure \ref{fig2}(a),  we show $d(t)$ as a function of $t$.
This aymmetric oscillation has turned out to be
important in order to induce gliding motion.
This is the reason why simple harmonic motion cannot induce gliding motion.
Due to this asymmetry,
as $d(t)$ becomes larger, velocity of oscillation becomes higher,
in contrast to the simple harmonic motion.
When oscillation is asymmetric as shown above,
$\theta(t)$ can resonate to the oscillation
(Figure \ref{fig2}(b))
if $\omega_0$ is chosen properly.
This means that the period of fluttering is forced close to
that of oscillation.
Moreover, as shown in Figure \ref{fig2}(c),
this resonance violates time reversal symmetry of $\theta(t)$
which $\theta(t)$ for periodic fluttering has.
Because of this symmetry breaking, periodic fluttering motion
becomes gliding motion although fluttering motion does not
vanish completely (See Figure \ref{fig3}).
\begin{figure}
\includegraphics[width=80mm]{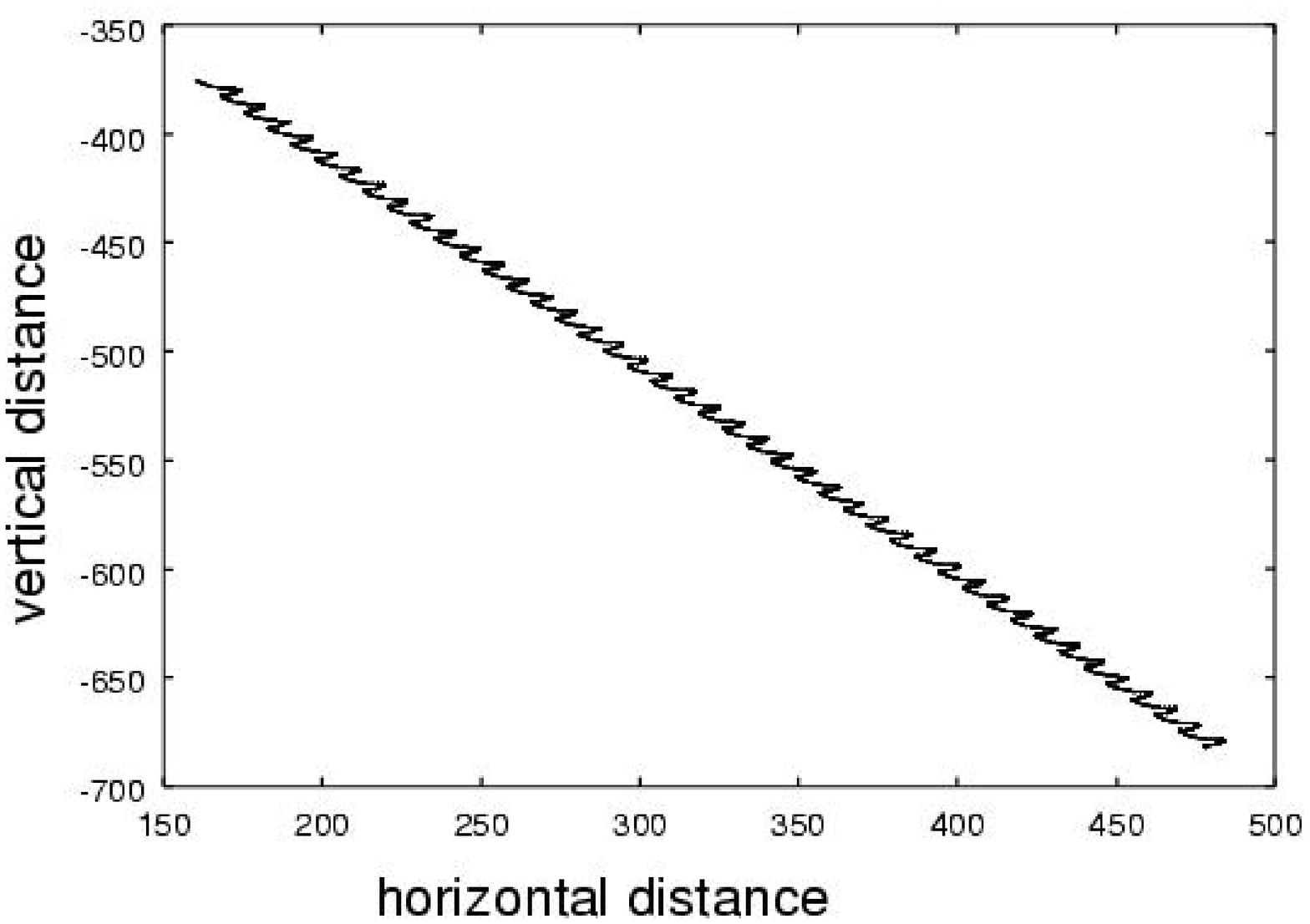}
\includegraphics[width=80mm]{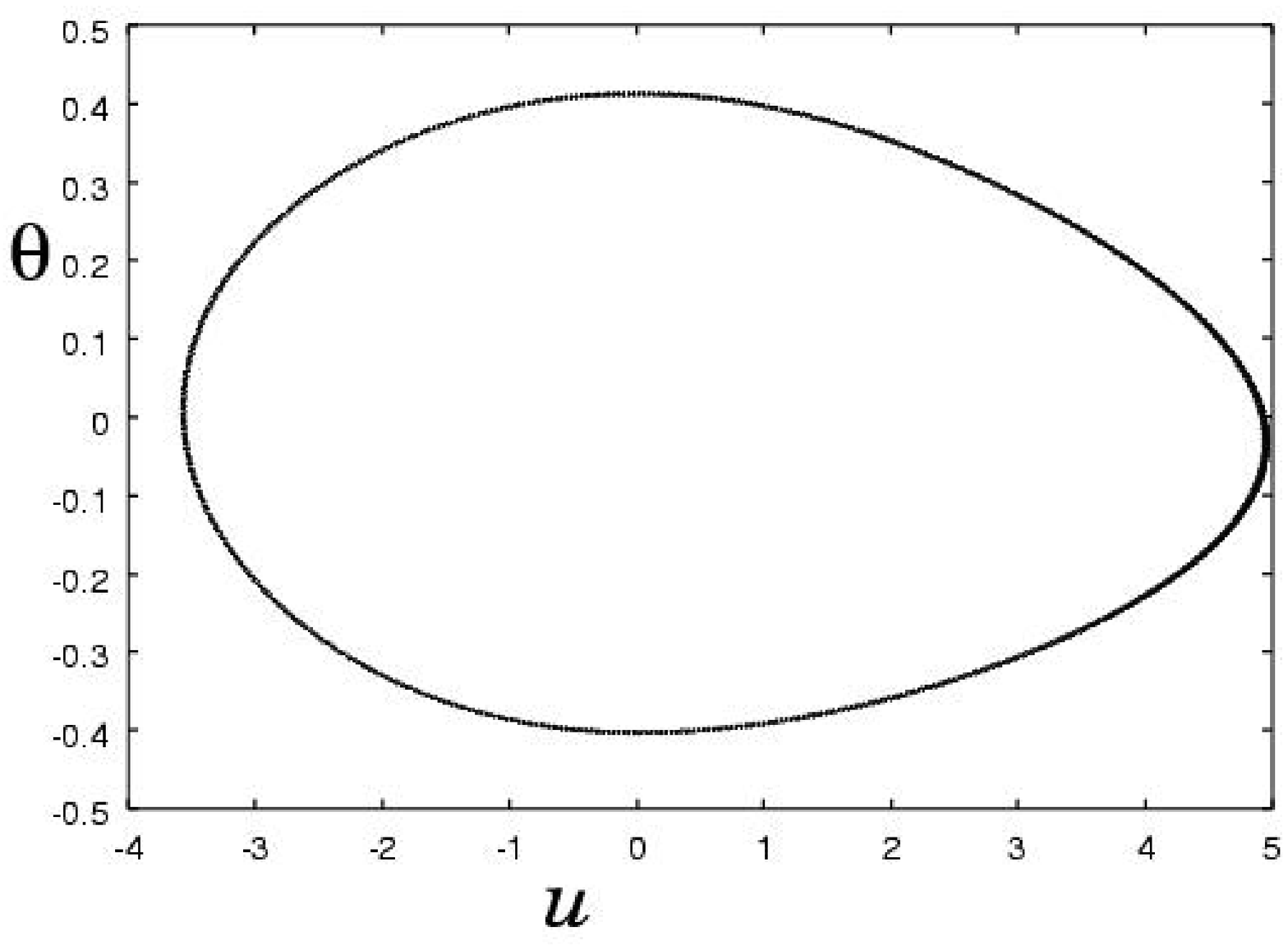}
\caption{Left: A trajectory in vertical plane. Right:
$\theta(t)$ vs $u(t)$ for the left figure.
The symmetry seen in the motion without oscillation (Figure \ref{fig1}) has
vanishes.}
\label{fig3}
\end{figure}
For a long period, a falling paper keeps constant velocity and constant angle
$\theta(t)$ as if flying snake keeps them\cite{snake}.
Also in Figure \ref{fig3}, we show $\theta(t)$ as a function of $u(t)$.
Since asymmetric motion of $\theta(t)$ results in that of
$u(t)$, in average a falling paper can move along one direction.
This is the reason why oscillation causes gliding motion, i.e.
a falling along one direction.

\begin{figure}
\includegraphics[width=54mm]{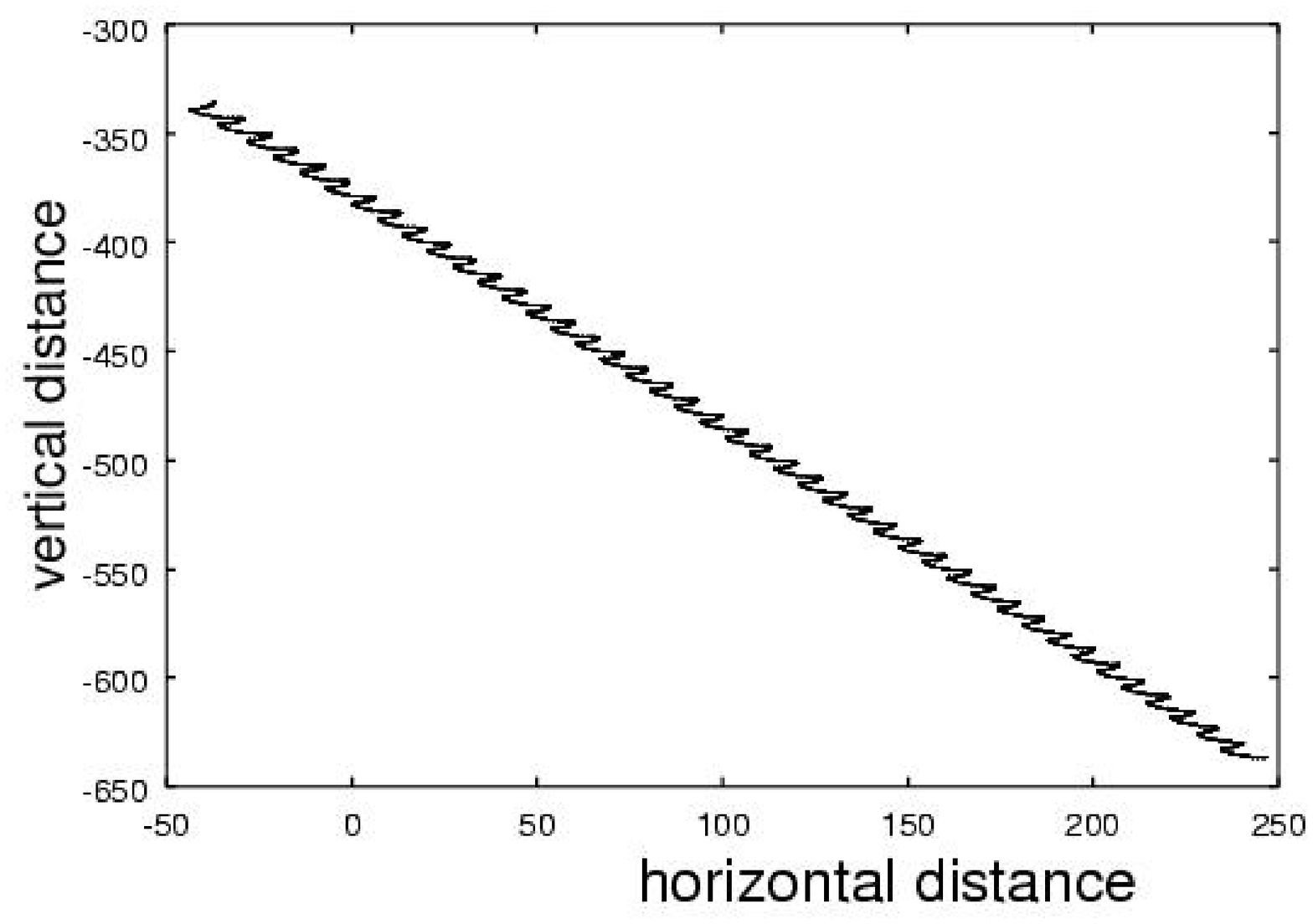}
\includegraphics[width=54mm]{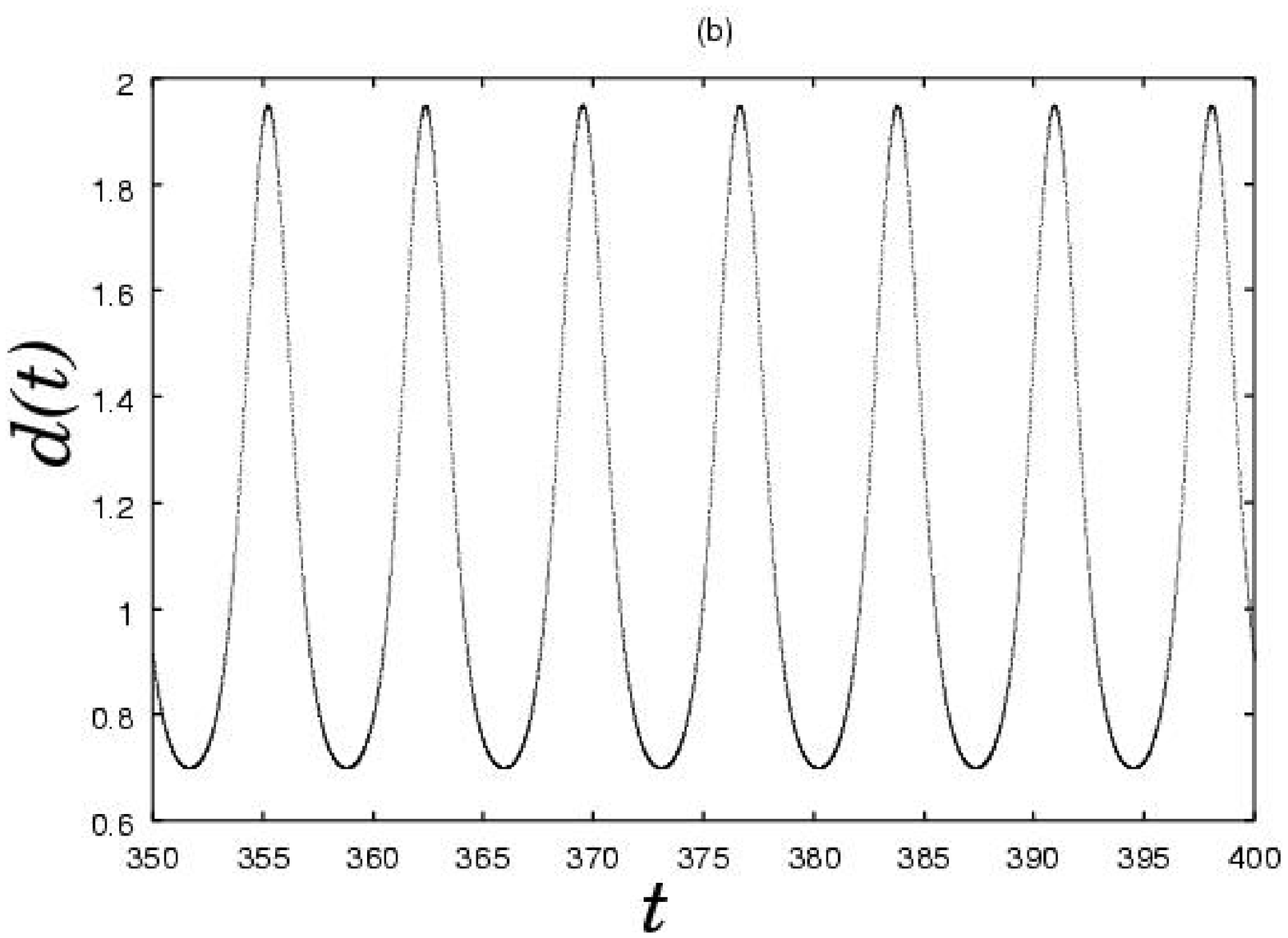}
\includegraphics[width=54mm]{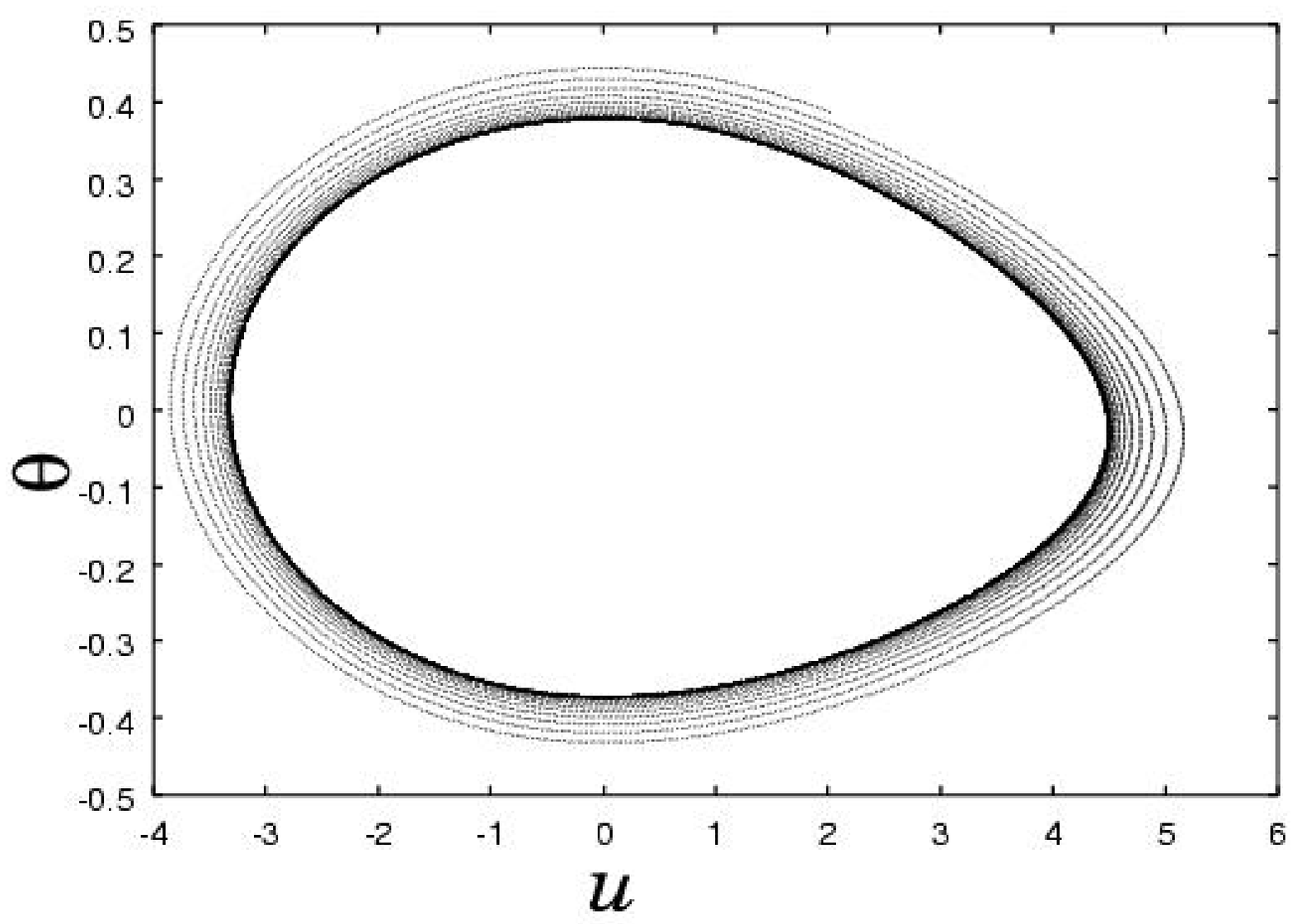}
\caption{Left:$d(t)=\exp(-1-1.3 \sin\omega_0 t)$, $\omega_0=0.9$ 
Center:  A trajectory in vertical plane. 
Right: $\theta(t)$ vs $u(t)$ for the center figure.
}
\label{fig4}
\end{figure}
\begin{figure}
\includegraphics[width=54mm]{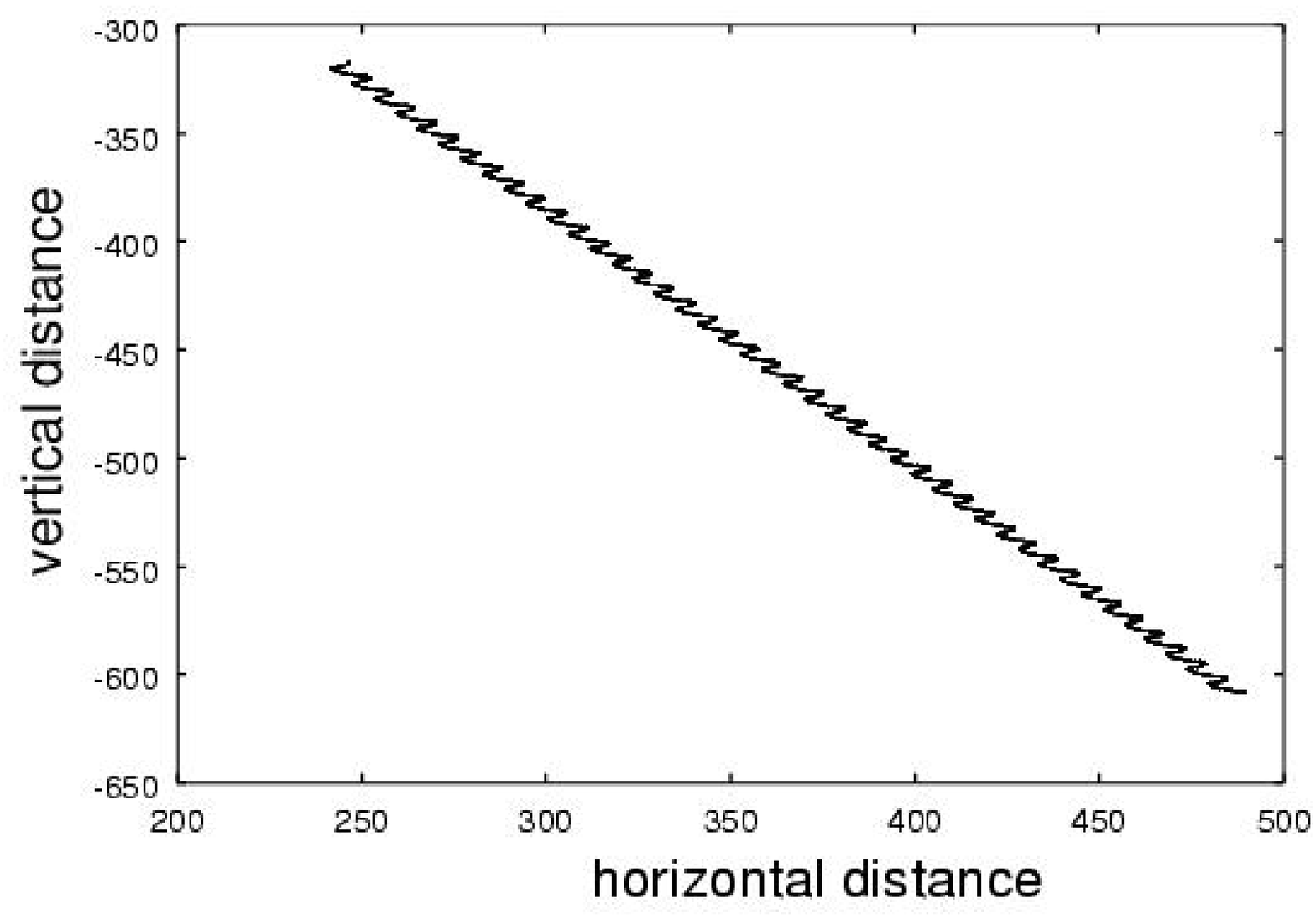}
\includegraphics[width=54mm]{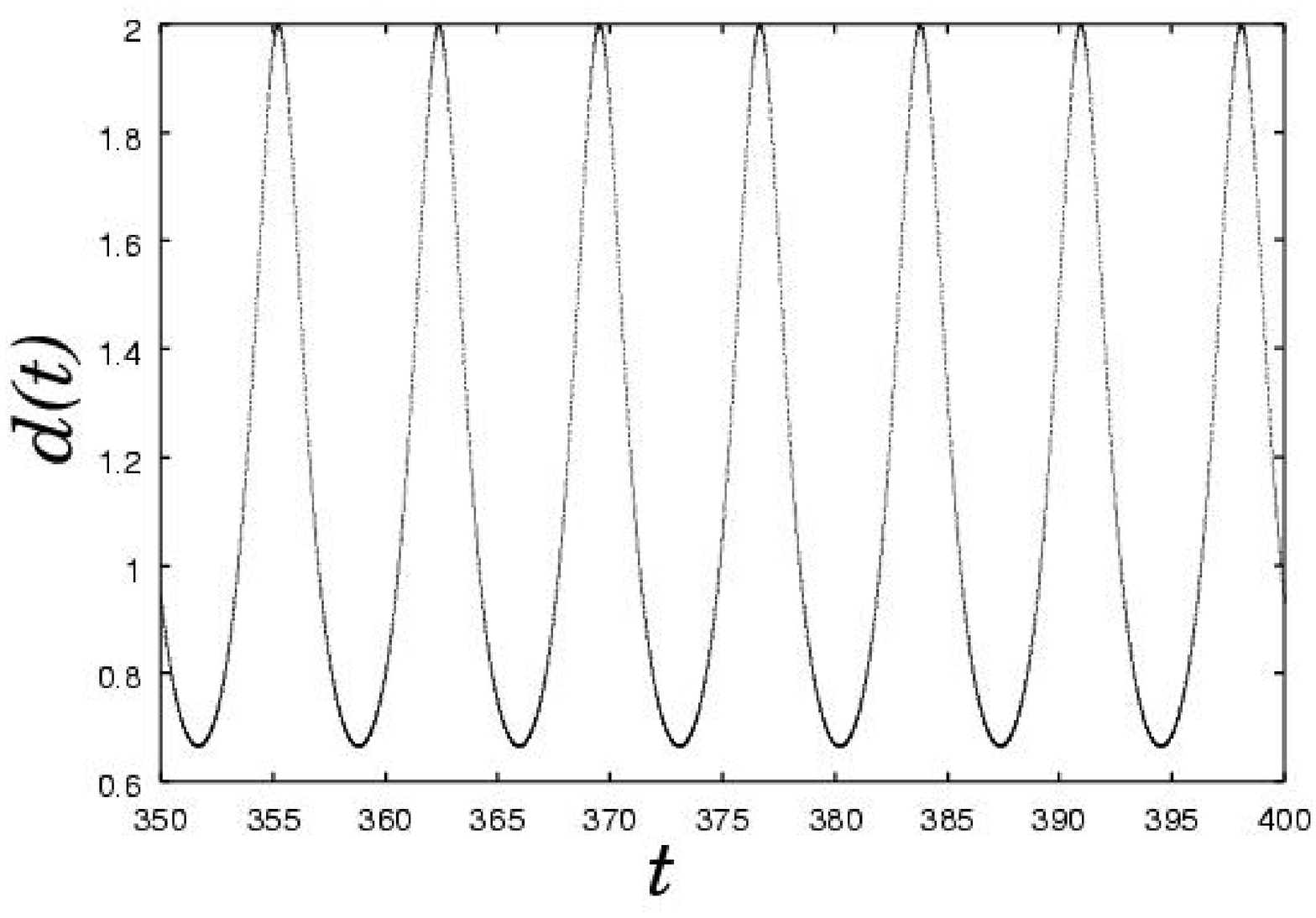}
\includegraphics[width=54mm]{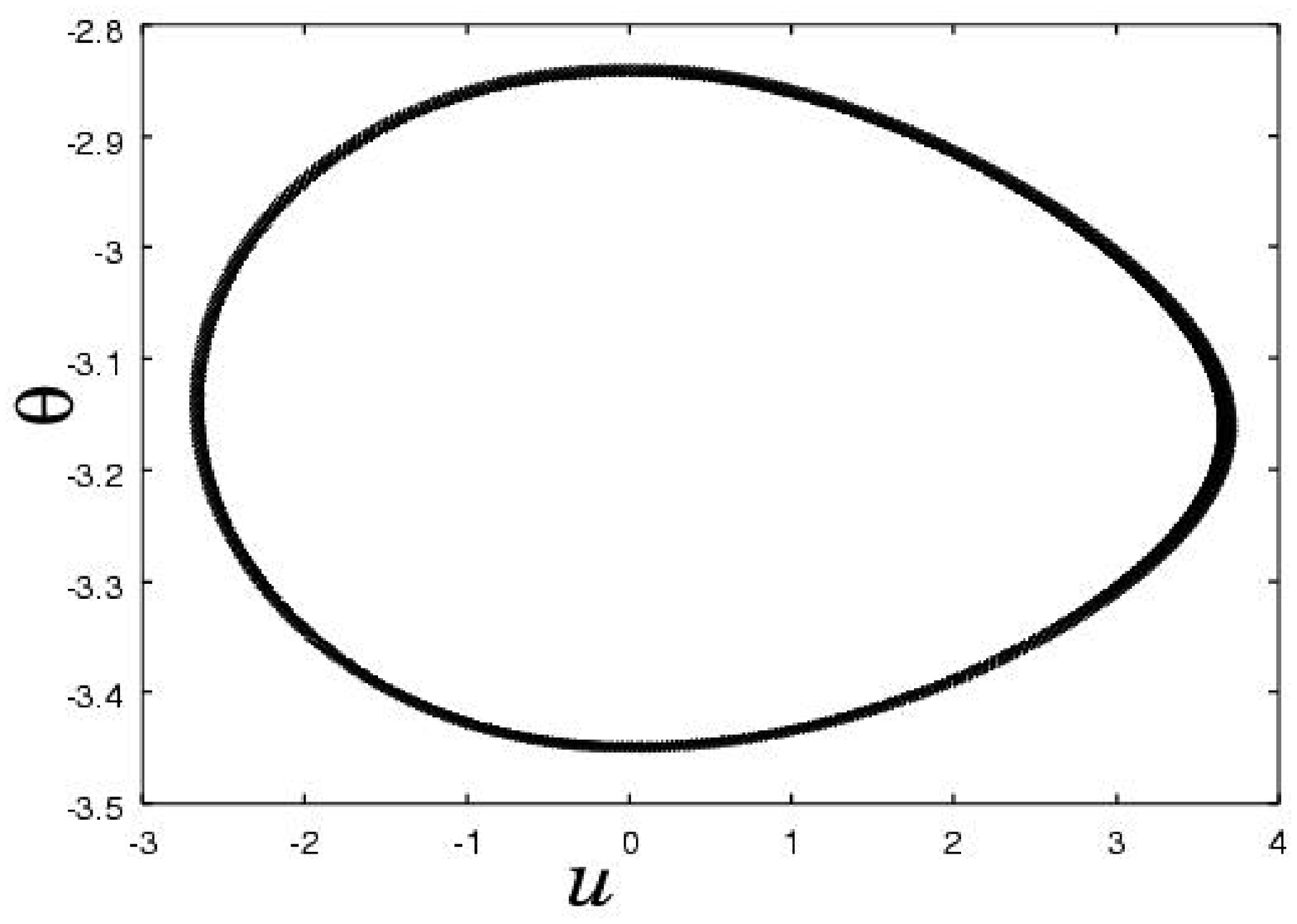}
\caption{Left:$d(t)=1/[1+0.5\sin\omega_0 t] $, $\omega_0=0.9$ 
Center:  A trajectory in vertical plane. 
Right: $\theta(t)$ vs $u(t)$ for the center figure.
}
\label{fig5}
\end{figure}
For the remaining two cases, we also show  trajectory, $d(t)$, 
and $\theta(t)$ vs $u(t)$ (Figures \ref{fig4} and \ref{fig5}).
In order to induce gliding motion,
for these two cases, $d(t)$ must
have time dependencies similar to that of the first case, too.
Although all of them induces the motions
along the one direction,
there are some differences among these three cases.
For the first case, $\theta - u$ plot shows very narrow band, which means
gliding motion is very periodic.
On the other hand, for the second case, $\theta - u$ plot
tends to converge to the periodic motion very slowly.
And for the third case, $\theta - u$ plot shows relatively
broader band, this means simple periodic 
motion is modulated by
slower periodic motion.

In spite of these minor differences, their global behaviours are very similar. 
Thus, there should exist
much more such possibilities of $d(t)$ that cause
gliding motion. 
Of course, $d(t)$ cannot take so different shapes from 
the above cases.
As long as we tried, only these shapes can induce gliding motion.
Probably, a flying snake would have found
such very rare possibilities
during its evolutions, because it has much more time for trial
than we spent. 

It is also a problem that we cannot stop
fluttering motion completely.
However, a flying snake can choose any other much more complicated time
dependence of $d(t)$
Further detailed investigation of
snake's motion will tell us which $d(t)$ can induce a
simple gliding motion that
is not accompanied with the fluttering.

\section{Sensitivities to the detailed parameter values or the forms of the
function $d(t)$}

Finally, we would like to comment on
sensitivities to the detailed parameter values or the forms of the
function $d(t)$.
Once we fix all parameter values other than those included in $d(t)$, 
we can change the functional form of $d(t)$ very little.
For example, if we slightly change the value $0.3$ in the form of
$d(t)=0.3(1+\sin \omega_0 t)^{2}+0.7$, 
the directional gliding motion disappears quickly.
Actually speaking, three $d(t)$s mentioned above are
tuned so that they have almost equal time dependency.
Thus, we believe that this specific form is very important.

\begin{figure}
\rotatebox{-90}{\includegraphics[height=54mm]{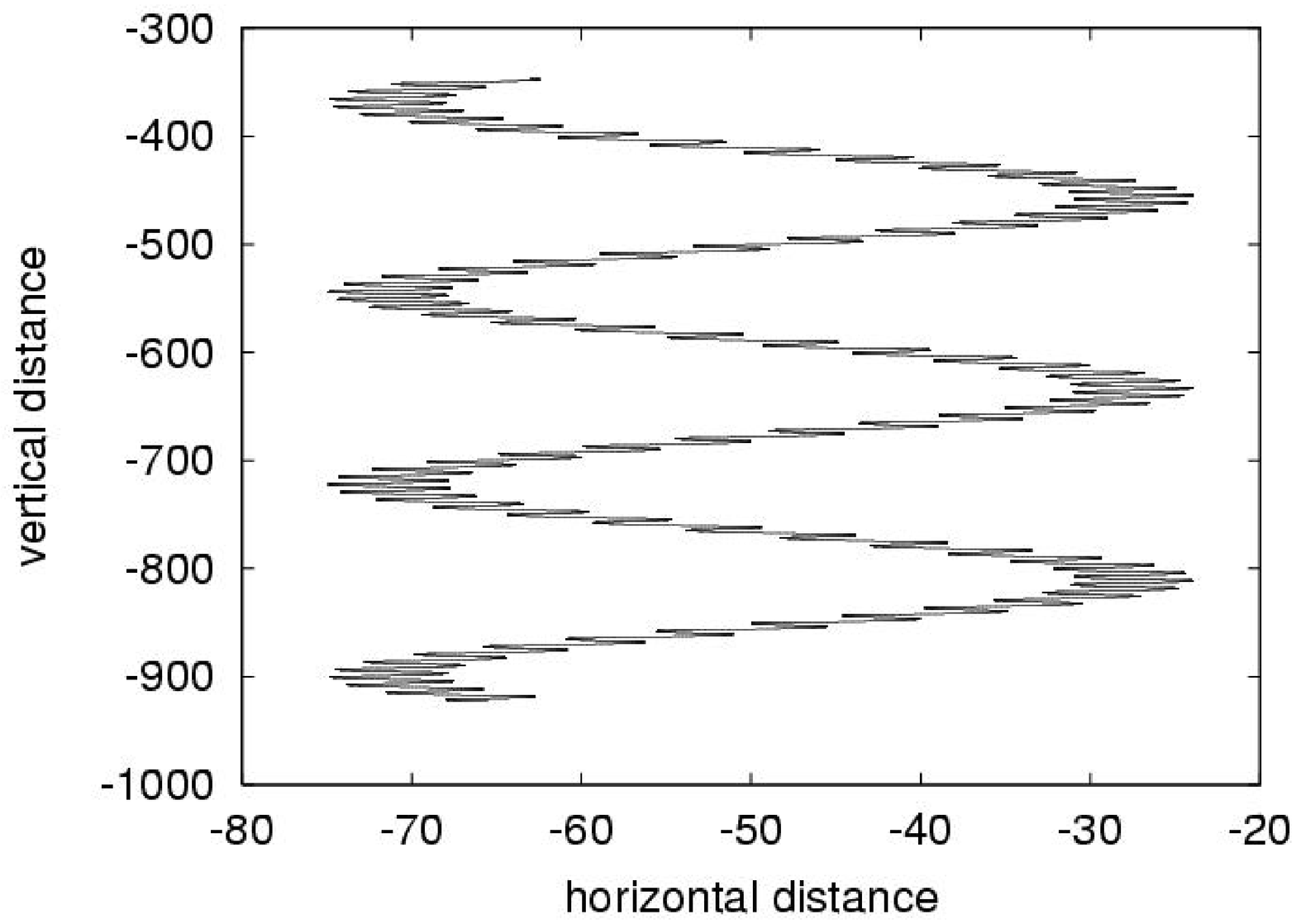}}
\rotatebox{-90}{\includegraphics[height=54mm]{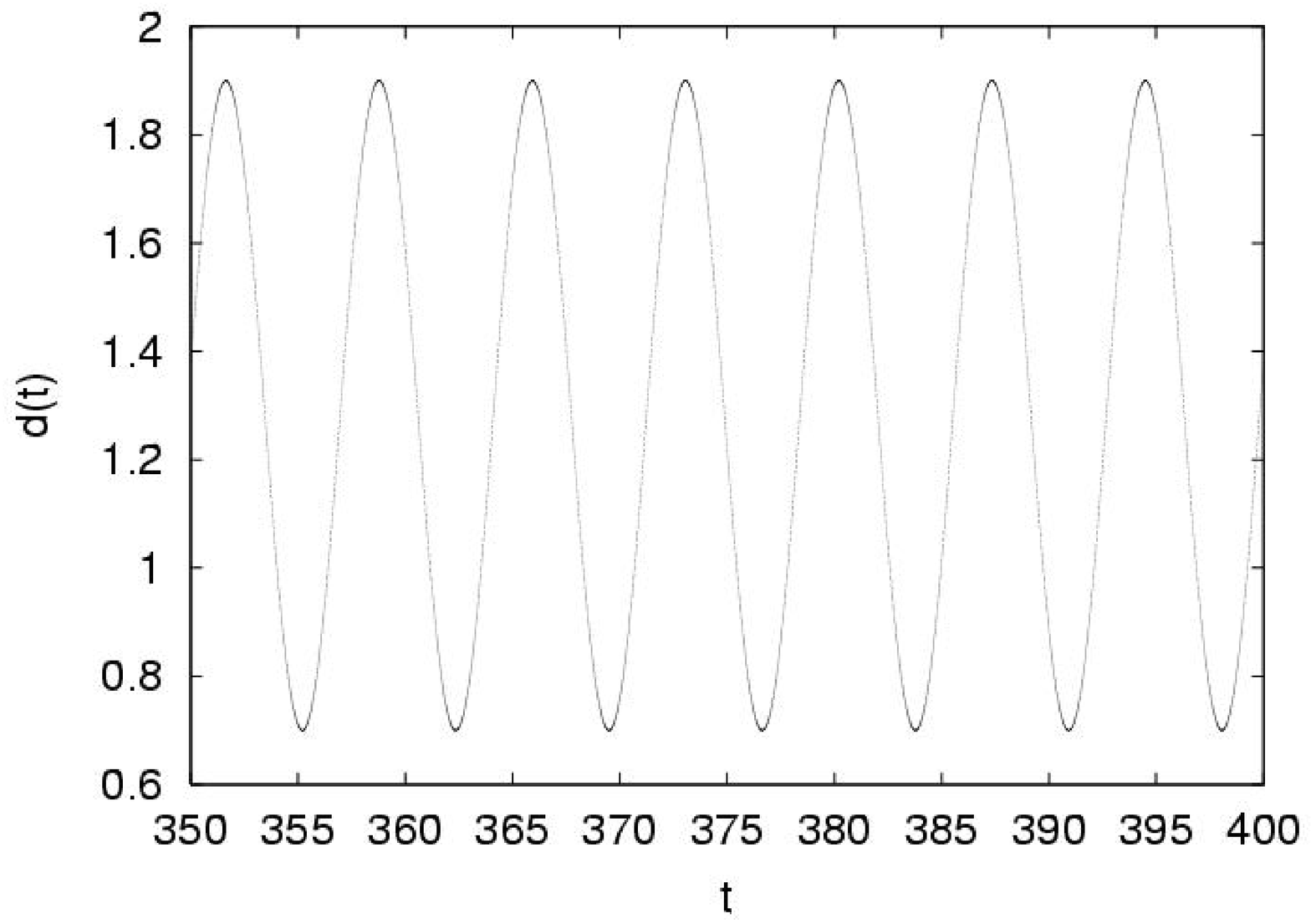}}
\rotatebox{-90}{\includegraphics[height=54mm]{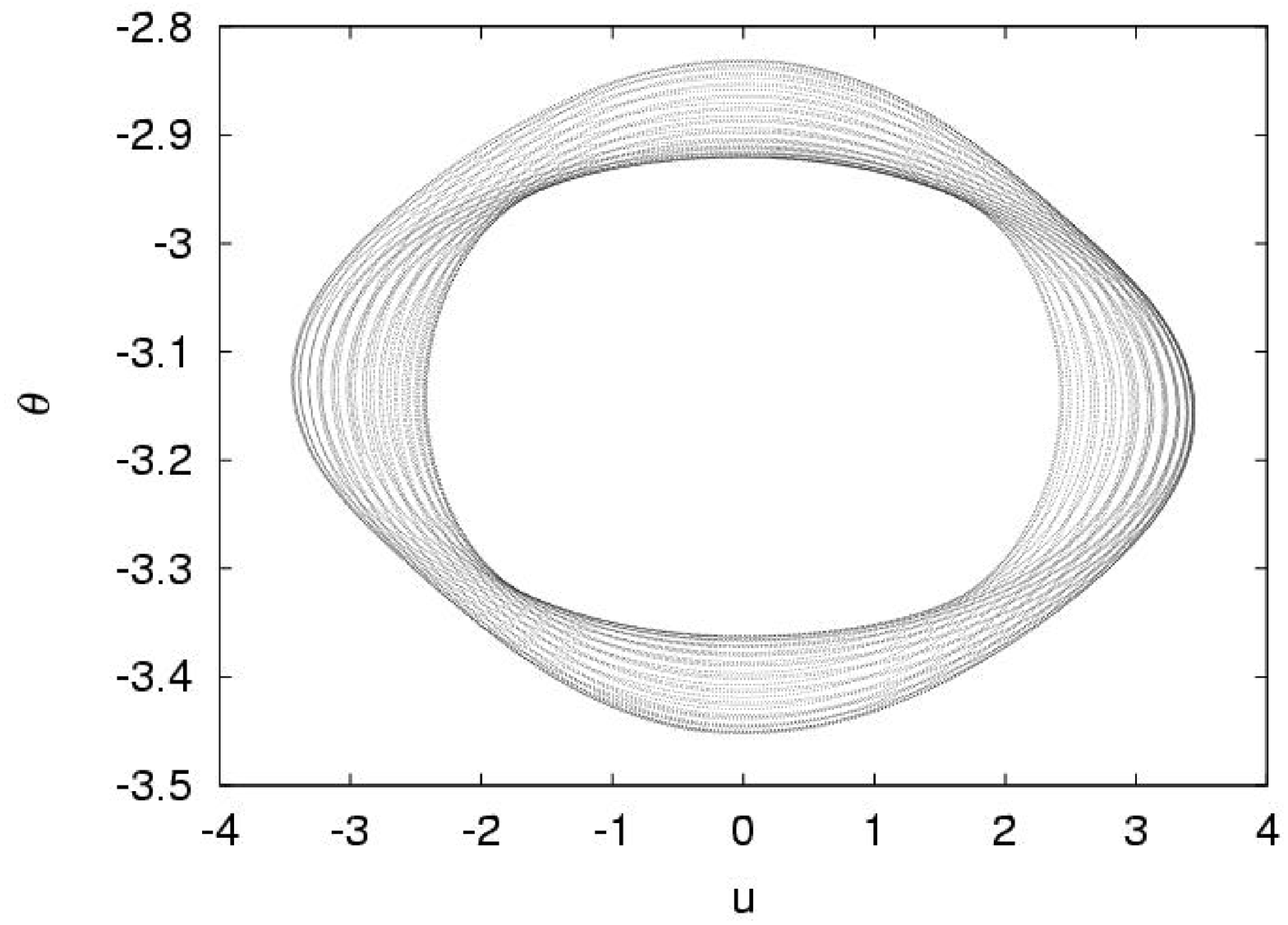}}
\caption{Left:$d(t)=0.6 \sin\omega_0 t+1.3 $, $\omega_0=0.9$ 
Center:  A trajectory in vertical plane. 
Right: $\theta(t)$ vs $u(t)$ for the center figure.
}
\label{fig6}
\end{figure}

Also, the condition for the directional gliding motion
is asymmetry. As mentioned above, asymmetry in $d(t)$ results in
that in velocity which causes directional motions.
For example, if we employ a normal sinusoidal motion
$$
d(t)=0.6 \sin\omega_0 t+1.3, \omega_0=0.9,
$$
there is not any asymmetry in velocity, thus directional gliding motions
cannot appear (Fig. \ref{fig6}).
In spite of the fact that $d(t)$ is tuned so that its functional form is close to those of three $d(t)$s
excluding the asymmetry, directional motion cannot occur.
Thus it is clear that asymmetry is the most important factor.

It is also easily understood that the direction of motion is
determined by the initial condition. Asymmetry in $d(t)$ cannot
decide in which direction symmetry is broken. This depends upon
how the snake starts to fly.


\begin{thebibliography}{99}
\bibitem{num}We solve the equation of motion
(\ref{eq1}) using the forth order Runge-Kutta method.
Parameters used are as follows:
$k_\perp = 10$, $k_\parallel=k_\perp/80$ and $\rho =0.1$. 
Initial conditions are $\omega=0,\theta=1, u=2,v=1$.
Under these conditions, original model  (i.e. $d(t)=1$)
induces periodic flattering motion.
\end{thebibliography}
\end{document}